\documentclass{winnower}
\usepackage{graphicx}
\usepackage{float}
\usepackage{graphicx,caption}

\usepackage{hyperref}
\definecolor{darkblue}{rgb}{0, 0, 0.5}
\hypersetup{colorlinks,urlcolor=darkblue,citecolor={darkblue}}
\usepackage[symbol]{footmisc}
\usepackage[all]{nowidow}
\usepackage{natbib}

\setcitestyle{authoryear}
\setcitestyle{citesep={;}}
\AtBeginDocument{}
\begin{document}

\makeatletter
\makeatother

\newcommand\blfootnote[1]{%
  \begingroup
  \renewcommand\thefootnote{}\footnote{#1}%
  \addtocounter{footnote}{-1}%
  \endgroup
}

\renewcommand{\thefootnote}{\arabic{footnote}}


\title{\noindent\rule[.25cm]{\textwidth}{1.5pt} Can you hear me \textit{now}? \\ Sensitive comparisons of human and machine perception \noindent\rule{\textwidth}{1.5pt} \vspace{.2cm} 
}

\author[ ]{Michael A. Lepori \{\href{mailto:mlepori1@jhu.edu}{mlepori1@jhu.edu}\}}

\author[ ]{Chaz Firestone \{\href{mailto:chaz@jhu.edu}{chaz@jhu.edu}\}}


\affil[ ]{Johns Hopkins University}

\date{}

\maketitle

\blfootnote{
\\ Department of Psychological \& Brain Sciences\\Johns Hopkins University\\3400 N Charles St\\Baltimore, MD 21218\\\\
Version: 6/27/22; in press, \textit{Cognitive Science}}

\vspace{-1.2cm}

\renewcommand{\abstractname}{\vspace{-\baselineskip}}
\begin{abstract}

\vspace{.3cm}

\noindent \textbf{The rise of machine-learning systems that process sensory input has brought with it a rise in 
comparisons between human and machine perception. But such comparisons face a challenge: Whereas machine perception of some stimulus can often be probed through direct and explicit measures, much of human perceptual knowledge is latent, incomplete, or unavailable for explicit report. Here, we explore how this asymmetry can cause such comparisons to misestimate the overlap in human and machine perception. As a case study, we consider human perception of \textit{adversarial speech} --- synthetic audio commands that are recognized as valid messages by automated speech-recognition systems but that human listeners reportedly hear as meaningless noise. In five experiments, we adapt task designs from the human psychophysics literature to show that even when subjects 
cannot freely transcribe such speech commands (the previous benchmark for human understanding), they often 
can demonstrate other forms of understanding, including discriminating adversarial speech from closely matched non-speech (Experiments 1--2), finishing common phrases begun in adversarial speech (Experiments 3--4), and solving simple math problems posed in adversarial speech (Experiment 5) --- even for stimuli previously described as unintelligible to human listeners. 
We recommend the adoption of such 
``sensitive tests'' when comparing human and machine perception, and we discuss the broader consequences of such approaches for assessing the overlap between systems. 
}\\

\end{abstract}

\section{Introduction}

How can we know when a machine and a human perceive a stimulus the same way? Machine-recognition systems increasingly rival human performance on a wide array of tasks and applications, such as classifying images \citep{Krizhevsky:2012wl, szegedy2016rethinking}, transcribing speech \citep{chan2016listen,lamere2003cmu}, and diagnosing mechanical or biological anomalies \citep{cha2017deep,Lakhani:2017ep}, 
at least on certain established benchmarks (for critical discussion, see \citealp{raji2021ai}). Such advances often call for 
comparisons between human and machine perception, in which researchers collect responses from human subjects and machine-recognition systems and then ask how similar or different those responses are. In some cases, these comparisons serve to establish standards for determining when a machine-recognition system has reached so-called ``human-level'' performance (e.g., by recording average human accuracy rates on visual and auditory classification tasks). In other cases, the purpose of such comparisons is subtler, as in work that 
uncovers aspects of human perception that are shared with various machine-learning systems \citep{Schrimpf:2018dp}. For example, recent work asks whether humans and machines that demonstrate comparable 
overall accuracy nevertheless show different patterns of errors \citep{Eckstein:2017gz,Rajalingham:2018kv}, exhibit different biases \citep{Baker:2018jp,buolamwini:2018gs}, or are susceptible to different illusions and ``attacks'' \citep{Elsayed:2018ni,jacob2019deep,ward2019exploring,Zhou:2019jxa}. 

\subsection{Knowing more than you can say}


Regardless of their goals, any comparison between human and machine perception must confront several challenges associated with the different constraints 
these systems face. 
One such challenge involves finding the right tasks and measures to serve as the basis of 
these comparisons, since different tasks may be better suited to assessing comprehension in humans vs.\ in machines. Indeed, whereas machine-recognition systems are often evaluated by more direct and explicit measures (since their classification decisions and other outputs are openly available to an experimenter), such measures  --- especially paradigms involving free descriptions of some stimulus under minimal constraints --- are typically understood to be inadequate tests of human perception. One reason for this is that human perception almost always involves incomplete knowledge, thresholded responses, and even unconscious processing that may not be cognitively accessible to the perceiver, such that assessing what someone knows or perceives is rarely as straightforward as asking them to describe, label, or categorize what they see, hear, or feel.

Indeed, decades of research on human perception and cognition have revealed knowledge and abilities that were not initially evident from tasks in which subjects freely report what they know or observe. For example, even when subjects cannot explicitly predict how a swinging ball will travel when released from a pendular trajectory, they may accurately place a cup to 
catch the ball --- suggesting that they possessed the relevant physical knowledge all along but were able to access it only through more implicit processes \citep{Smith:2018ff}. Similarly, even when subjects incorrectly report the locations of remembered objects, they may perform reliably above chance if asked to take a second or third guess (\citealp{Wu:2018iu}; see also \citealp{Vul:2008cz}). Even when subjects report no awareness of objects that are masked or appear outside the focus of attention, they may nevertheless show priming effects for the unnoticed stimuli, suggesting that those stimuli were processed unconsciously or below the threshold for explicit report (\citealp{kouider2007levels,Mack:2003hn}; for critical discussion, see \citealp{phillips2018unconscious}). Even when subjects fail to notice statistical regularities in visual displays, they may still apply those regularities on subsequent trials, implying that they learned and incorporated such regularities implicitly \citep{chun2000contextual,chun1998contextual}. And in an especially dramatic case, patients with cortical blindness who fail to freely report features of the objects they are looking at (e.g., being unable to answer questions like \textit{``what orientation is the line in front of you?''}) may nevertheless succeed under forced-choice conditions (e.g., being able to correctly answer questions like \textit{``is the line in front of you horizontal or vertical?''}; \citealp{weiskrantz1986blindsight,weiskrantz1996blindsight}; also \citealp{phillips2021blindsight}).  

\subsection{\textit{Sensitive tests} in human-machine comparisons}

The above examples involve what we will refer to here as \textit{sensitive tests} of human perception and cognition. Sensitive tests are tasks or measures that go beyond simply asking someone to describe what they see, hear, or know. Such techniques include making subjects 
act on a piece of information (rather than report it), exploring downstream consequences for other behaviors (as in priming studies), collecting additional responses (such as ranking various options rather than giving a single answer), or using some piece of knowledge to make a 
discrimination (rather than trying to report that knowledge directly). 

How might this apply to comparisons between humans and machines? The fact that human perceptual knowledge can be partial, incomplete, or buried beneath layers of unconscious mental processing creates a challenge for comparisons of human and machine perception. In particular, whenever such comparisons rely mostly or only on explicit descriptions of sensory stimuli, there is a risk that these measures may underestimate what the human subjects really know about the stimuli they perceive, and thereby misestimate the overlap between human and machine perceptual processing.\footnote{Whether similar considerations also apply to tests of 
machine perceptual knowledge is interesting open question. For relevant work on this issue, see \cite{Zoran_2015_ICCV,Ritter:2017uk}. For a more general discussion of similar themes, see \cite{firestone2020performance}.}


Here, we suggest that these considerations 
matter in concrete and measurable ways. In particular, we argue that some apparent gaps or disconnects between human perceptual processing and the processing of various machine-perception systems can be explained in part by insufficiently sensitive tests of human perceptual knowledge. To demonstrate this, we explore an empirical case study of how using more sensitive tests can reveal a perceptual 
similarity when previous studies seemed to show a deep 
dissimilarity. 
Accordingly, we recommend that comparisons of human and machine perception adopt such sensitive tests before drawing conclusions about how their perceptual processing differs.


\subsection{A case study: Adversarial misclassification}

An especially striking difference in human and machine perception is the one implied 
 by adversarial misclassification \citep{szegedy2014intriguing}. Adversarial examples are inputs designed to cause high-confidence misclassifications in machine-recognition systems, and they may crudely be divided into two types. The first type is sometimes called a ``fooling'' example, in which a stimulus that would otherwise be classified as meaningless or nonsensical (e.g., patterns of image static, or auditory noise) is recognized as a familiar or valid input by a machine (e.g., a dog, or the phrase ``OK Google, take a picture''; \citealp{Nguyen:2015cv,Carlini:vl}). The second type is a ``perturbed'' example, in which a stimulus that would normally be classified in 
one way (e.g., as an orange, or a piece of music) can be very slightly altered to make a machine classify it in a completely 
different way (e.g., as a missile, or the command  ``Call 911 now'') --- even when such perturbations seem irrelevant (or are not even noticeable) to human observers \citep{Athalye:2018ml,szegedy2014intriguing}.

Such misclassifications are significant for at least two kinds of reasons. First, and more practically, they expose a major vulnerability in the security of machine-perception systems: If machines can be made to misclassify stimuli in ways that humans would not notice, then it may be possible to 
attack such systems in their applied settings (e.g., causing an autonomous vehicle to misread a traffic sign, or making a smartphone navigate to a dangerous website) --- a worry that may only intensify as such technologies become more widely adopted \citep{Hutson:2018ks}. Second, and more theoretically, two systems classifying the same stimulus so differently would seem to undermine any other similarities they might show \citep{brendel2020adversarial}, and even rule out the use of one to model the other.  

Crucially, the 
reason adversarial misclassifications carry such important and interesting consequences is the very strong and intuitive sense that machines 
perceive these stimuli in ways that humans 
do not. And indeed, a growing literature has sought to demonstrate this empirically, by asking human subjects to classify such stimuli and noting similarities and differences in their classification decisions (\citealp{Carlini:vl,chandrasekaran2017takes,Elsayed:2018ni,Harding:2018vx,yuan2020adversarial, Zhou:2019jxa}; see also \citealp{Baker:2018jp,dujmovic2020adversarial,feather2019metamers,golan2019controversial}). 
But might some of these discrepancies arise in part because of the means of comparison themselves?

\subsubsection{Can people understand adversarial speech?}

\begin{figure*}
\begin{center}
\vspace{.2cm}
\includegraphics[scale=0.20]{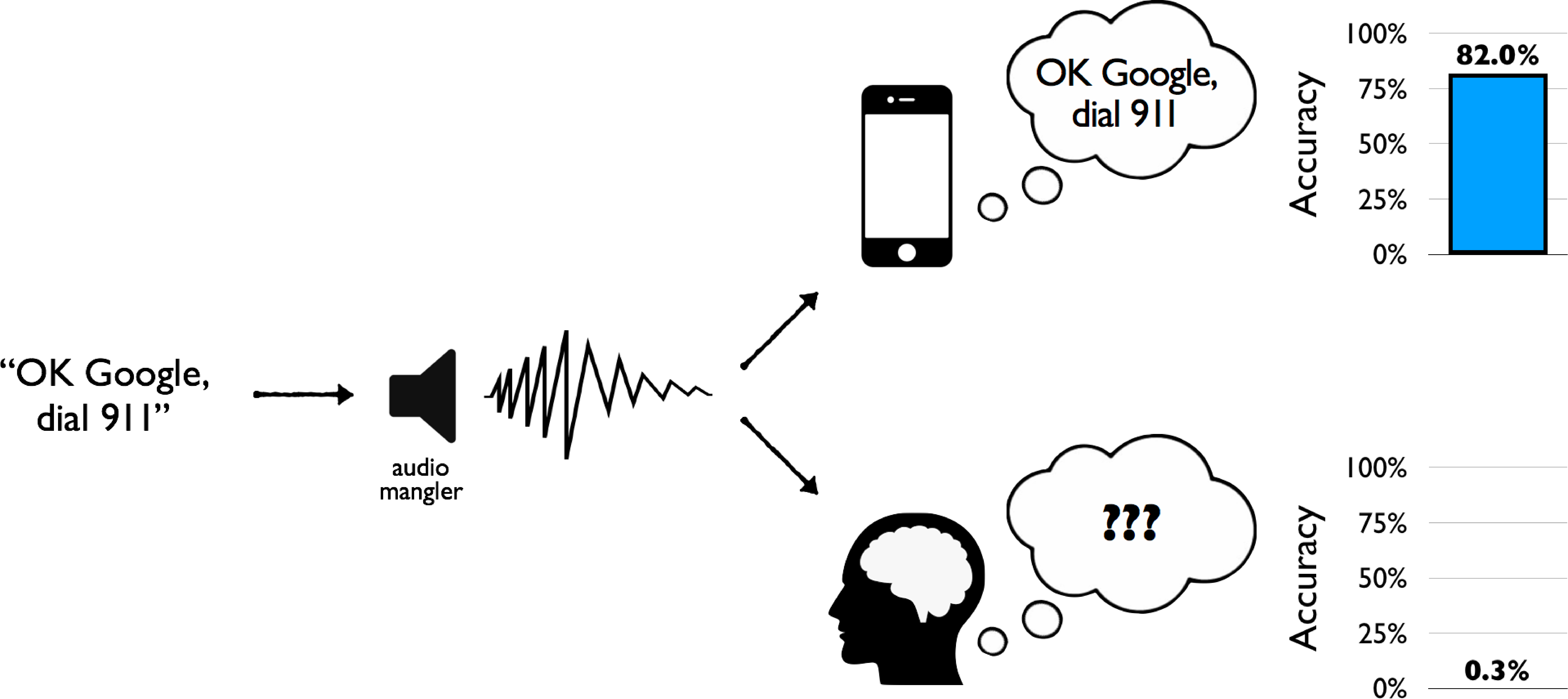}
\captionsetup{width=.9\linewidth,font=footnotesize}
\vspace{.3cm}
\caption
{\textit{Adversarial audio commands}. Whereas human listeners and automated speech-recognition systems may transcribe normal human speech with comparable accuracy, adversarial speech involves synthetic or modified audio clips that human listeners hear as meaningless noise but that are recognized as valid commands by automated speech-recognition systems. As shown in \cite{Carlini:vl}, it is possible to take a speech clip and pass it through an ``audio mangler'' that erases many features that humans rely on in order to understand speech. This procedure results in a signal that is reliably recognized as a valid command by the machine-recognition system targeted by the attack, while human listeners are reportedly unable to understand it (as measured by their ability to freely transcribe its contents).}
\label{fig:mangler}
\end{center}
\end{figure*}

As a case study of this possibility, we consider here the example of 
adversarial speech. A recent and influential research program shows that it is possible to generate audio signals that are recognized as familiar voice commands by automated speech-recognition systems but that human listeners hear as meaningless noise (\citealp{Carlini:vl}; Figure \ref{fig:mangler}). In short (though see below for more detail), a normal voice command can be ``mangled'' by removing audio features not used by the speech-recognition system, such that it remains perfectly intelligible to that system but becomes incomprehensible to a human listener\footnote{According to some restrictive definitions of ``adversarial examples'', stimuli only get to count as adversarial if they very closely resemble the original stimuli from which they were created. By those standards, the present case involves a misclassification, but perhaps not an adversarial one. By contrast, here we assume the broader and more popular definition given by Goodfellow et al. (\citealp{goodfellow_papernot_huang_duan_abbeel_clark_2017}; see also \citealp{szegedy2014intriguing}): ``Adversarial examples are inputs to machine learning models that an attacker has intentionally designed to cause the model to make a mistake''. By this definition, the case we explore here easily fits the bill.}. To verify that such stimuli are truly heard as meaningless non-speech, this work included an empirical comparison between the relevant speech-recognition system and a cohort of human subjects, who were played the audio clips and asked to make judgments about them. In particular, the comparison asked the subjects to transcribe the audio files, and found that no subjects were able to transcribe the adversarial speech clips into the underlying messages from which the files were produced. These and similar tests led the authors to conclude that ``0\%'' of subjects heard the hidden messages in the files, that subjects ``believed our audio was not speech'', and more generally that these commands were ``unintelligible to human listeners''.

However, accurate transcription is an almost paradigmatically ``insensitive'' test --- an extremely high bar to study comprehension of this sort, including for the practical and theoretical issues raised above. For example, even in an applied setting, it could be valuable to know whether a human listener can tell 
that a hidden command was just played, even if the listener does not know the precise nature of the command (so that, e.g., the user could monitor or disable their device if they suspect it is being attacked). Similarly, for at least some purposes, understanding even 
part of a message might be nearly as good as understanding the whole thing. For example, a smartphone user who heard mostly nonsense in an adversarial speech clip --- but managed to pick up the words ``9-1-1'' --- might understand what their phone is about to do, even if they could not make out all or even most of the full message (e.g., ``OK Google, please dial 911''). This is perhaps especially likely if they have a sense of the adversary's intentions or probable messages (e.g., the kinds of commands one would give to a phone in the first place). Finally, beyond such applied concerns, discovering that humans could extract these subtler patterns of meaning from adversarial speech would imply greater processing overlap (or least fail to imply a lack of overlap) than previous tests seemed to reveal. 
Might more sensitive tests show that this is the case?

\subsection{The present experiments: Sensitive tests of adversarial speech comprehension}

Here, we use perhaps the simplest of such sensitive tests --- forced-choice classification --- to explore how differences between human and machine perception may be overestimated by insensitive tests of human perceptual knowledge. We generate adversarial speech commands using the same method as in \cite{Carlini:vl}. But rather than ask subjects to freely transcribe these messages, we probe their understanding by asking them to make discriminations 
based on the commands. 
Experiments 1 and 2 ask whether subjects can discriminate adversarial messages that contain English speech from messages that do not contain speech but that are otherwise closely matched. 
Experiments 3 and 4 ask whether subjects can supply the last word in a familiar phrase begun in adversarial speech (e.g., ``\textit{It is no use crying over spilled $\rule{0.6cm}{0.15mm}$}''), even without any advance knowledge of what words will be played. Finally, Experiment 5 asks whether subjects can answer simple math problems posed in adversarial speech (e.g., ``$2 + 5 =\,\, ?$''). To foreshadow the key results, all five experiments show that subjects can demonstrate comprehension of such speech clips when tested under sensitive conditions, even though these clips are not easily transcribed.

Of course, this work covers only one approach to generating adversarial speech (see the General Discussion for a review of other approaches), and adversarial speech is itself only one example of apparent divergence between human and machine perception. Our intention is thus for the particular results we report below to serve as a case study of a more general lesson: Sensitive tests can reveal latent or incomplete perceptual knowledge that less sensitive tests can miss, and in ways that directly inform comparisons between human and machine perception.

\vspace{.5cm}
\begin{center}
\section{General Methods}
\end{center}
\vspace{.2cm}


Though the five experiments reported here involve different hypotheses, designs, and research questions, all five proceeded in a very similar way and shared many methodological characteristics. What follows is thus a general methods section applying to all of the experiments below, followed by shorter methods sections for each individual experiment covering those key factors that differed.   

\vspace{.5cm}
\subsection{Open Sciences Practices}

The hypotheses, experimental designs, analysis plans, exclusion criteria, and sample sizes were determined in advance --- and formally pre-registered --- for all five experiments. The data, materials, code, and pre-registrations for all experiments reported here are available at \url{https://osf.io/kp5j9/}. 

\subsection{Participants}

Each experiment recruited 100 
different
subjects (i.e., 500 subjects total across all five experiments) from Amazon Mechanical Turk (\url{https://mturk.com}). For discussion of this subject pool's reliability, see \cite{Crump:2013fn}, who use this platform to replicate several classic findings from cognitive and perceptual psychology. All subjects consented to participate in the study and were monetarily compensated for their participation. 

All experiments applied exclusion criteria that required successful performance on attention-check trials and tests of the subjects' audio quality. These criteria varied slightly across experiments, and were always pre-registered. Across all experiments, an average of 77\% of subjects passed all exclusion criteria and were thus included in subsequent analyses. However, we note that no result reported here depended in any way on these exclusions; in other words, all of the results reported below remained statistically significant, in the same direction, even without excluding any subjects at all.

\subsection{Generating Audio Commands}

All experiments involved the presentation of adversarial speech. 
\citet{Carlini:vl} present two methods for generating these speech stimuli: a black-box method, which is agnostic to the targeted speech recognition system, and a white-box method, which requires prior knowledge of the particular speech system of interest.
To generate these speech stimuli, we used the white-box method.\footnote{We thank Nicholas Carlini for generously sharing the code to create these stimuli.}
We chose the white-box method because it was identified by the original authors as ``significantly better than the black-box attack'' at fooling human subjects. For this attack (and only for this attack), it was claimed that ``0\%'' of subjects understood the speech clips, and even that subjects ``believed our audio was not speech''. 
The white-box method is designed to fool the CMU Sphinx speech recognition system.\footnote{Note that CMU Sphinx \citep{lamere2003cmu} has a non-neural architecture (based on a Hidden Markov Model) and so is in many ways 
unlike the systems that are much discussed today as sharing deeper aspects of human perception and cognition. But this fact only serves to strengthen any positive results in our experiments: If we can demonstrate greater-than-expected overlap between human perception and a machine-perception system that is 
not typically thought to have architectural similarities to the human mind and brain, then it should seem all the more impressive if humans 
are able to understand such messages. (Moreover, it is not actually clear that HMMs and other non-neural architectures are necessarily `worse' as models of human perception and cognition; indeed, they were regularly used for just that purpose in a previous generation of computational cognitive science; \citealp{kaplan2008overview,miller1952finite}.)} In brief, the Spinx system extracts features from the raw audio input using the Mel-Frequency Cepstrum transformation (Mel-Frequency Cepstrum Coefficients; MFCCs), as well as simple transformations on the computed values. These features are used as input to a Hidden Markov Model (HMM), whose states correspond to phonemes. This HMM outputs a probability distribution over feature vectors that could be generated next. A Gaussian Mixture Model (GMM) is used to represent this distribution over feature vectors.

The white-box method exploits knowledge of the parameters the GMM (coefficients, means, and standard deviations for each Gaussian), as well as the dictionary used to transform words into phonemes. Given a string of text, this algorithm decomposes that string into a sequence of phonemes, and then attempts to generate input audio that will cause the Sphinx's HMM to generate that sequence. To do this, the algorithm uses knowledge of the GMM's parameters in order to calculate a series of MFCC feature vectors that maximize the likelihood under the GMM. Finally, the white-box method uses gradient descent in order to generate audio clips that are optimized to produce these MFCC feature vectors. Throughout this process, the audio corresponding to each phoneme is kept as short as possible, which further distorts the final clip.

\citet{Carlini:vl} notes that Sphinx correctly transcribes the audio clips generated by this method when they are fed directly into the system as audio files, but they are sometimes not transcribed correctly when they are played through a speaker and recorded with a microphone. They describe a method to improve the adversarial audio's over-the-air transcription ability, but we choose to use stimuli that do not correct for these difficulties. We do this to maximize the difficulty of comprehending these hidden audio commands, so that it will be especially clear how sensitive tests can reveal such comprehension.

In a supporting study (run after all of the studies reported below), we confirmed that the adversarial speech stimuli are inaudible when measured by free transcription. We selected one clip from each of the studies we report below, and showed 100 subjects either adversarial or uncorrupted text-to-speech versions of these stimuli. These subjects were tasked with transcribing the audio clip
s, as in \citet{Carlini:vl}. 
 
We ran two versions of this task: One in which subjects could play the clip
s only once, and one which (a separate group of 100) subjects could play the clips as often as they liked. In the single-play condition, subjects in the adversarial condition transcribed the audio correct
ly 0.6\% of time, and transcribed the uncorrupted stimuli correctly 97\% of the time. In the multi-play condition, subjects in the adversarial condition performed at 0\% (failing to transcribe even a single example correctly), whereas subjects in the uncorrupted condition transcribed the stimuli correctly 99.1\% of the time. This assured us that our adversarial stimuli were as inaudible as \citet{Carlini:vl}'s (who report similar performance).

Finally, we note that the approach we take here (which uses very same code as in the original study) builds into the stimulus-generation process all of the internal knowledge of the particular version of Sphinx that was targeted by the original, 
and so essentially `fools' that internal model of Sphinx as it generates audio.

All audio files used here, as well as the code for generating them, are available in our archive of data and materials.




\vspace{.5cm}
\begin{center}
\section{Experiment 1: Which One is Speech?}
\end{center}
\vspace{.2cm}
\label{exp1}
Can more sensitive tests reveal deeper human understanding of adversarial speech? Experiment 1 first asked whether forced-choice conditions could allow human subjects to distinguish adversarial speech from closely matched non-speech (even without requiring that subjects report the content of the speech; see Experiments 3--5 for tests of such contentful comprehension). We synthesized several dozen adversarial speech commands that previous work suggested should be ``unintelligible to human listeners'' and even ``believed [to be] not speech'' \citep{Carlini:vl}, and then played these commands to subjects either 
forwards or 
backwards (Figure \ref{fig:exp1}A). We predicted that subjects would hear the forwards-played audio as more speech-like than the backwards-played audio, even though the two kinds of clips were matched on 
many low-level features
(since these two trial types involved the very same audio clips --- the only difference was whether they were played forwards or backwards). If so, this would suggest that subjects do hear such audio clips 
as speech after all, in ways that would suggest a greater overlap in how such audio is processed by human listeners and the relevant speech-recognition systems. 

\vspace{.5cm}
\subsection{Methods}
\subsubsection{Stimuli}

We generated 54 hidden audio commands using the procedure described above. To select the content of the speech commands, we chose common idioms, quotes from history and media, or natural sequences --- for example, ``laughter is the best medicine'', ``we have nothing to fear but fear itself'', and ``1 2 3 4 5 6 7 8 9 10 11 12''. (See materials archive for the full list of phrases.) We chose such phrases instead of completely arbitrary collections of words so that we could roughly match the familiarity of the phrases used in \cite{Carlini:vl} --- which included, for example, the phrases ``take a picture'' and ``text 12345''. Thus, in both our study and in past studies, the stimuli included words and phrases that a typical listener would have heard before, even though the subjects had no advance knowledge of which 
particular words would appear. Notably, this is also the case for the likely words that a malicious attacker might transmit to a smartphone or home assistant (e.g., messages involving key words or phrases such as ``call'', ``browse'', ``unlock'', etc., as well as small numbers).

In addition to the 54 audio clips containing these messages, we also generated a corresponding set of 54 audio files that simply played those very same clips in 
reverse. This process ensured that these two sets of stimuli were minimal pairs, 
matched for many low-level auditory characteristics, including average length, frequency, intensity, and so on (as well as the variance in such characteristics). Thus, these pairs differed primarily 
in whether they followed the patterns characteristic of human speech (though see \citealp{irino1996temporal}). 

Finally, we generated one audio file containing a simple tone, to be used as a ``catch'' trial to ensure that subjects were engaged in the task and paying attention (see below). This file appeared 5 times in the experiment.

There were thus 109 audio files (54 Speech + 54 Non-Speech + 1 Catch), and 59 trials (54 experimental trials containing one forwards clip and its backwards counterpart, and 5 catch trials). All clips were generated, stored, and played in .wav format.

\subsubsection{Procedure} 

Subjects were told a brief story to motivate the experiment:

\begin{quote}
    \textit{A robot has hidden English messages in some of the following audio clips. Can you help us figure out which clips contain the robot's messages? We know that half of the messages have hidden English and half of them don't; we want your help figuring out which ones are English and which are not.}
\end{quote}

The experiment proceeded in a self-paced manner, with subjects triggering the playing of the clips. On each trial, subjects completed a two-alternative forced-choice task (2AFC). Two embedded audio players appeared onscreen, each loaded with a single clip that was played when the subject hit a ``play'' button. The two clips were always forwards and backwards versions of the same adversarial audio command (with left-right position on the display always randomized for each trial). After the subject played each clip at least once, they could select whether the left or right clip was the one that contained English speech. Subjects could play each clip additional times if they chose to before responding. After making their selection, the next trial began and proceeded in the same way. The command played on each trial was always randomly chosen (without replacement) from the 59 total trials (54 Experimental and 5 Catch), each of which was played for every subject.

Note that, even though these clips were of fairly well-known phrases, subjects had no advance knowledge of the particular words they should look for in the clips. As has been noted previously \citep{Carlini:vl}, adversarial speech can sometimes be easier to decipher when one knows (or is ``primed'' by) what the message is supposed to be; but, as in previous work, no such knowledge or priming was possible here (beyond knowledge that the message might be drawn from the extremely broad class of messages that includes all vaguely familiar phrases, idioms, and sequences in English). 

To ensure that subjects were paying attention and that the audio interface was working properly, subjects were instructed about how to behave on Catch trials: ``A few times in the experiment, instead of hearing some sounds from the robot, you will instead hear a simple beep or tone; whenever that happens, make sure to click `Right'.'' No data from Catch trials was included in our analysis, except as criteria for exclusion. Additionally, before beginning the experiment, a ``sound test'' was performed in which a single audio clip was played (Bach's \textit{Cello Suite No. 1}), and subjects had to say what kind of audio the clip contained (beeping, clapping, conversation, traffic, music, or ocean). Only if subjects selected ``music'' could they proceed to the experimental trials; any other option ended the experiment without collecting data. 

We excluded subjects based on two criteria. First, any subject who failed to provide a complete dataset was excluded from our analysis. Second, any subject who failed to follow instructions on any one of Catch trials (i.e., who selected ``Left'' when they should have selected ``Right'') was excluded entirely. This was done to ensure that all subjects had read the instructions and were focused on the experiment. These exclusion criteria, along with the rest of the design and analysis plan, were pre-registered.

Finally, as an additional sanity check, we also repeated the above experimental design using the 5 original adversarial speech stimuli created by \citet{Carlini:vl}, so that we could compare our results to the original work (for details, see Appendix \ref{carliniStimRep}).

Readers can experience this task for themselves at \url{ https://perceptionresearch.org/adversarialSpeech/E1}.

\begin{figure*}[!h]
\begin{center}
\vspace{.2cm}
\includegraphics[width=.6\linewidth]{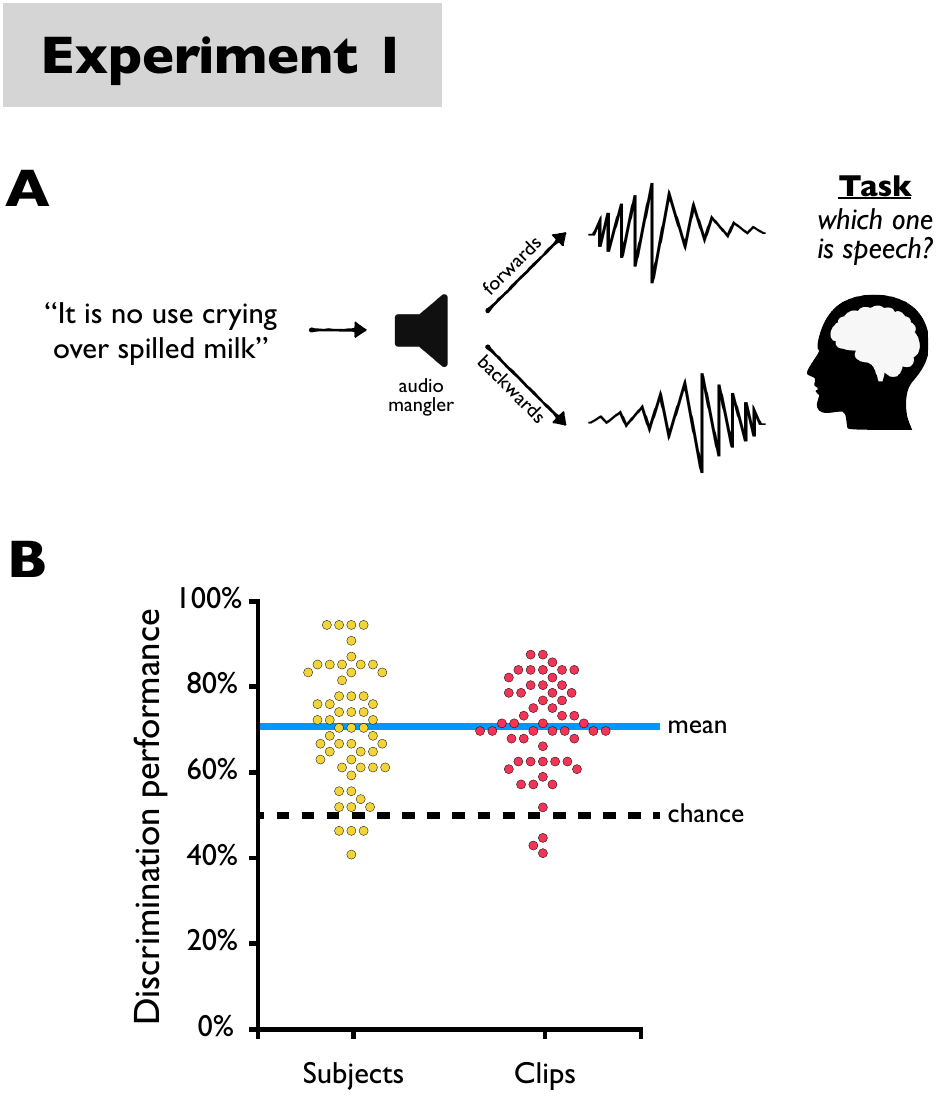}
\captionsetup{width=.9\linewidth,font=footnotesize}
\vspace{.1cm}
\caption{
\textit{Design and Results of Experiment 1}. (A) On each trial, subjects heard two adversarial speech clips, one forwards-played and one backwards-played version of the same adversarial audio command; their task was just to say which contained English speech. (B) Subjects correctly discriminated adversarial speech from these closely matched non-speech clips. The vast majority of subjects classified the clips correctly more often than they classified them incorrectly (yellow), and the vast majority of speech clips were classified correctly more often than they were classified incorrectly (red).}
\label{fig:exp1}
\end{center}
\end{figure*}

\subsection{Results and Discussion} 

Subjects demonstrated an ability to discriminate adversarial speech from closely matched non-speech. Average accuracy on experimental trials was 70.6\%, which significantly differed from chance accuracy (50\%), $t(55) = 11.34, p < .001, d = 1.529$ (Figure \ref{fig:exp1}B, blue)\footnote{A direct replication of Experiment 1, with extra reminders for how to behave on catch trials, excluded fewer subjects and produced a similar pattern of results: 63.8\%, $t(72) = 8.43, p < .001, d=.994$.}. Thus, subjects were able to reliably determine whether an adversarial audio clip contained speech.

In addition to the pre-registered analysis we report above, we also conducted two exploratory analyses whose logic closely followed previous work \citep{Zhou:2019jxa}. Beyond the ``raw'' accuracy across all subjects and all trials (i.e., the proportion of 
trials in which subjects correctly classified a clip as Speech or Non-Speech), it may also be informative to know (a) the proportion of \textit{subjects} who tended to classify such stimuli correctly (collapsing across all speech clips), as well as (b) the proportion of 
speech clips that tended to be classified correctly (collapsing across all subjects). 

In fact, collapsing across all speech clips, 92.9\% of subjects showed classification performance that was numerically above chance (Figure \ref{fig:exp1}B, yellow). In other words, the vast majority of subjects tended to classify such clips correctly rather than incorrectly, suggesting that the ability to hear adversarial speech as speech was quite widespread. Additionally, collapsing across all subjects, 94.4\% of the 54 speech clips were classified correctly more often than they were classified incorrectly (Figure \ref{fig:exp1}B, red).\footnote{Note that this does not mean that each of these subjects performed 
significantly above chance, or that each of these clips were identified as speech significantly above chance --- only that 92.9\% of subjects got the right answer more often than they got the wrong answer (whereas chance performance would predict that only 50\% of subjects would do so), and that 94.4\% of speech clips were classified correctly more often than they were classified incorrectly (whereas chance performance would predict that only 50\% of clips would be correctly identified in this way). As stated in the main text, mean performance on any given trial was 70.6\%.}

Finally, we also discovered a strikingly similar pattern of results when this paradigm was repeated with \citet{Carlini:vl}'s own stimuli. When subjects heard the 5 original adversarial speech clips studied in past work, they performed at 68.9\% discriminating speech from non-speech, which significantly differed from chance accuracy (50\%), $t(82) = 8.20, p < .001, d = 0.906$ (for additional detail, see Appendix \ref{carliniStimRep}).

In other words, whereas previous methods involving speech transcription suggested that ``0\%'' of subjects heard the hidden messages in the files, or that subjects ``believed our audio was not speech'', our approach here suggested that a large majority of subjects heard the speech clips as speech more often that not, and that nearly all clips were heard as speech more often than not. These results thus provided initial evidence that a more sensitive approach (here, using forced-choice classification) could reveal perceptual knowledge and abilities that less sensitive tests were unable to detect.\footnote{See Appendix~\ref{carliniStimRep} for a replication of this experiment using stimuli from \citep{Carlini:vl}}

\vspace{.5cm}
\begin{center}
\section{Experiment 2: Speech or Non-Speech?}
\end{center}
\vspace{.2cm}

The previous experiment revealed an ability to discriminate adversarial speech from closely matched non-speech, when subjects are tested under sensitive forced-choice conditions. However, the psychophysically powerful 2AFC design may have given subjects an undue advantage, since success on this task requires only that subjects decide which clip 
sounds more speechlike, regardless of whether either of the clips actually 
sounds very much like speech. In other words, it is possible that subjects felt that 
neither clip sounded like speech much at all, but were still able to succeed on the test as long as they could tell which clip better resembled ordinary speech. Though this is, in many ways, the entire purpose of 2AFC task designs, people in the real world (i.e., where such adversarial attacks might actually be deployed) may not be in the position of subjects in Experiment 1. In that case, a natural question is whether subjects could identify adversarial speech as speech under looser conditions in which they must actively label a single clip as speech, rather than compare two clips.

Experiment 2 investigated this question by asking subjects to label single clips as speech or non-speech. The design was extremely similar to Experiment 1, except that instead of 54 experimental trials each containing two speech clips (one forwards and one backwards), there were 108 experimental trials each containing one clip (either the forwards or backwards version of the 54 audio commands). Subjects' task was now to classify each clip as speech or non-speech, rather than to decide which of two clips was more speechlike. Could subjects succeed even under these conditions? (Readers can experience this task for themselves at \url{https://perceptionresearch.org/adversarialSpeech/E2}.)

\vspace{.5cm}
\subsection{Results and Discussion} 

Subjects demonstrated an ability to identify adversarial speech clips as speech. Average accuracy on experimental trials was 62.2\%, which significantly differed from chance accuracy (50\%), $t(82) = 11.34, p < .001, d = 1.253$. Performance was comparable on forwards trials (63.0\%) and backwards trials (61.5\%). Thus, subjects were able to reliably determine whether an adversarial audio clip contained speech, by identifying adversarial speech as speech and closely matched non-speech as non-speech.

We also performed the same two exploratory analyses as described in Experiment 1. Collapsing across all speech clips, 91.0\% of subjects showed classification performance numerically above chance. And collapsing across all subjects, 84.3\% of the clips were classified correctly more often than they were answered incorrectly.

Thus, in addition to being able to tell the difference between adversarial speech and adversarial non-speech, subjects could also identify a given adversarial speech clip 
as speech.

\vspace{.5cm}
\begin{center}
\section{Experiment 3: Fill in the Blank}
\end{center}
\vspace{.2cm}

The previous experiments suggested that human subjects can identify and discriminate adversarial speech from closely matched non-speech. But this result says little or nothing about subjects' ability to understand the 
content of that speech. 
Indeed, it is possible that subjects were able to simply distinguish speech and non-speech using the low-level differences between normal and reversed audio clips. For example, natural audio has sharp attacks and long decays, which are not present in reversed audio \citep{irino1996temporal}.
Experiment 3 thus asked whether forced-choice conditions could allow human subjects to display knowledge of the content of adversarial speech, by asking them to identify the next word in an adversarial phrase.

We took a subset of the speech clips from Experiments 1--2 and simply removed the last word from the phrases (so that, e.g., ``\textit{It is no use crying over spilled milk}'' became ``\textit{It is no use crying over spilled}''), and then asked subjects to supply the final word under forced-response conditions. If they can do so, this would suggest that subjects can engage in deeper and more contentful processing of adversarial speech, beyond knowing which clips are speech and which are not. 

\vspace{.5cm}
\subsection{Methods}

Experiment 3 proceeded in the same way as Experiments 1--2, except as noted below.

\subsubsection{Stimuli}

To generate the stimuli, we analyzed performance on the 54 adversarial speech clips from Experiment 2, and selected the 20 phrases with the highest classification accuracy in that experiment. Using the same procedure described earlier, we generated new adversarial speech clips from these 20 phrases, shortened by one word. For example, ``laughter is the best medicine'' became ``laughter is the best''. (See materials archive for the full list of phrases.) Importantly, all of these missing ``last words'' (e.g., ``medicine'' above) were unique to a given adversarial speech clip, and none of these words were spoken in any of the other adversarial speech clips. These clips served as the experimental stimuli.

We also generated 3 files containing non-mangled speech using an online text-to-speech system. These commands contained an uncorrupted human voice reciting portions of the alphabet (e.g., ``A, B, C, D, E...''). These were used as catch trials to ensure that subjects were engaged in the task and paying attention.

\subsubsection{Procedure}

Subjects were told a modified version of the story from Experiments 1--2: 

\begin{quote}
    \textit{A robot has hidden English messages in some audio transmissions that we've recovered. However, these audio clips have been ``corrupted'', in two ways. First, most of the transmissions sound very strange and garbled; the robot's ``voice'' is very different than a human voice. Second, they've all been cut short by at least one word; for example, a message that was supposed to be ``O say can you see by the dawn's early light'' might actually come across as ``O say can you see by the dawn's early''.}
\end{quote}

After the subjects played the clip on a given trial (e.g., ``It is no use crying over spilled''), the two buttons that appeared either contained the correct next word in the current phrase (e.g., ``milk''), or a word that would complete a different experimental phrase (e.g., ``medicine''). The incorrect option was randomly selected from the pool of last words for other phrases in the experiment (without replacement), such that each last word appeared twice in the experiment, once as the correct option and once as the incorrect option. The pairs of options were randomly generated for each subject, as was the order in which the clips were shown. 
Note that, even though these clips were of fairly famous phrases, subjects had no advance knowledge of the particular kinds of words they should look for in the clips, just as in Experiments 1--2. The only property uniting these phrases was that they were likely to be vaguely familiar (rather than, say, being likely to be about food, sports, or some other theme).


As in previous experiments, we excluded any subject who failed to provide a complete dataset, as well as any subject who answered any of the Catch trials incorrectly.

Readers can experience this task for themselves at \url{https://perceptionresearch.org/adversarialSpeech/E3}.

\begin{figure*}[!h]
\begin{center}
\vspace{.2cm}
\includegraphics[width=.6\linewidth]{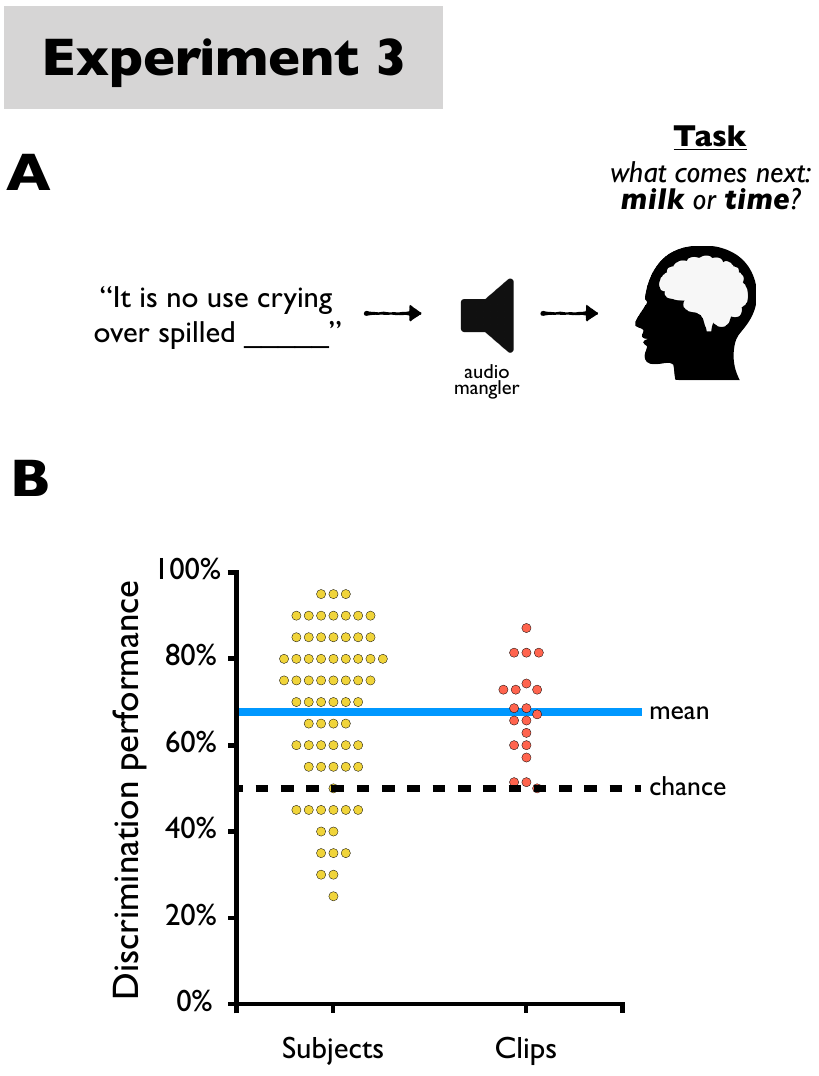}
\captionsetup{width=.9\linewidth,font=footnotesize}
\vspace{.1cm}
\caption
{\textit{Design and Results of Experiment 3}. (A) On each trial, subjects heard an adversarial speech clip that was missing its last word; their task was to identify the word that should come next. (B) Subjects correctly identified the next word in adversarial speech clips, and this was again broadly true across both subjects (yellow) and speech clips (red).}
\label{fig:exp3}
\end{center}
\end{figure*}

\subsection{Results and Discussion}

Subjects demonstrated an ability to identify the next word of a phrase contained in an adversarial speech command. Average accuracy on experimental trials was 67.6\%, which significantly differed from chance (50\%), $t(69) = 7.98, p < .001, d = 0.960$ (Figure \ref{fig:exp3}B, blue). Thus, subjects could reliably understand at least some of the content of the adversarial speech commands, and could apply this comprehension to identify the next word of the hidden phrase. Collapsing across all speech clips, 79.3\% of subjects showed classification performance numerically above chance (Figure \ref{fig:exp3}B, yellow). Thus, most subjects were able to correctly identify the next word in the hidden phrase. Collapsing across all subjects, 19 of the 20 clips were completed correctly more often than they were completed incorrectly (and 1 was completed correctly exactly half of the time; Figure \ref{fig:exp3}B, red).

These results suggest that subjects can not only hear adversarial speech 
as speech, but can also decipher at least some of the 
content of such speech. Whereas the task in Experiments 1--2 could have been solved by attending to phonetic or phonological features (such as prosody, intonation, or picking up one or two phonemes), Experiment 3 required some understanding at the level of 
words. This ability thus goes far beyond what was previously shown in transcription tasks, further showing how sensitive sensitive tests can reveal surprisingly robust perceptual knowledge in ways that less sensitive tests cannot.

\vspace{.5cm}
\begin{center}
\section{Experiment 4: Fill in the Blank, After a Single Play}
\end{center}
\vspace{.2cm}

Experiment 3 suggested that subjects can not only hear adversarial speech as speech, but can also comprehend the content of that speech. However, it is perhaps possible that the forced-choice options themselves assisted subjects in parsing the speech, and in a way that could potentially undermine our interpretation of that experiment. When one is given a clue about what to hear in obfuscated speech, it may be easier to hear that speech in line with one's expectations (suggesting that speech processing is subject to top-down influence; see especially \citealp{Remez:1981ub}, as well as discussion in \citealp{firestone2016cognition,firestone2016seeing,vinson2016perception}). In that case, one could imagine the following sequence of events: First, subjects heard an adversarial speech clip (e.g., ``laughter is the best $\rule{1cm}{0.15mm}$'') and 
initially found it completely incomprehensible; second, they noticed the possible answers (e.g., ``milk'' vs. ``medicine''); third, it occurred to them that not many well-known phrases end in ``medicine'', such that one of the possible phrases might be ``laughter is the best medicine'' (or ``a taste of your own medicine'', but perhaps few others); fourth, and finally, they re-played the adversarial speech clip while paying special attention to whether it might be the phrase ``laughter is the best'', and indeed were able to hear it that way.

Though this interpretation would still attribute to subjects 
some ability to comprehend adversarial speech, it is perhaps less impressive than subjects exhibiting this ability even without any such hints. So, to find evidence of this more impressive ability, Experiment 4 repeated Experiment 3 with two small changes. First, the ``last word'' options on a given trial were revealed only after the adversarial speech clip had completely finished playing, such that every subject first heard the clip without any keywords that might tell them what to attend to. Second, though we still gave subjects the ability to play the clips multiple times, we recorded the number of plays a given subject made on a given trial, and pre-registered a follow-up analysis including only those trials in which the subject chose not to re-play the clip (i.e., those trials on which the first, clue-less play was the only play). If subjects can still identify the missing word in a clip even without any hint in advance about which particular words might be in the clip, and even on only a single clue-less play of the clip, this would be especially compelling evidence that subjects can and do comprehend aspects of the 
content of adversarial speech.

Readers can experience this task for themselves at \url{https://perceptionresearch.org/adversarialSpeech/E4}.

\vspace{.5cm}
\subsection{Results and Discussion} 

Subjects again performed successfully. First, considering all trials (including multi-play trials), subjects anticipated the next word in the phrases with an accuracy rate of 67.8\%, which significantly differed from chance (50\%), $t(76) = 9.66, p < .001, d = 1.107$; this result replicates Experiment 3. But second, and more tellingly, even on trials in which subjects played the clip only a single time (such that they knew the last word options only 
after hearing the adversarial speech clip), subjects retained the ability to anticipate the next word in the adversarial speech phrases, with an accuracy rate on such trials of 69.8\%, which significantly differed from chance (50\%), $t(69) = 7.04, p < .001, d = 0.848$. Thus, subjects can comprehend the content of adversarial speech, even without advance knowledge of which words to hear in such clips.

\vspace{.5cm}
\begin{center}
\section{Experiment 5: Math Problems}
\end{center}
\vspace{.2cm}

Even though the results of Experiments 3--4 suggest that human listeners can hear meaning in adversarial speech commands, the task could still have been completed with a very minimal understanding. For example, a subject who was played ``It is no use crying over spilled'' could fail to comprehend almost the entire message, but still correctly guess ``milk'' over relevant foils if they just heard one telling keyword (e.g., ``crying''). Could subjects instead complete a task that required parsing an 
entire message spoken in adversarial speech? 

Experiment 5 asked whether subjects could correctly answer simple arithmetic problems that are contained in adversarial speech commands. We generated 20 audio commands containing these arithmetic problems (e.g., ``six minus two''), and then asked subjects to select the correct answer from two options. On one hand, this approach constrains the space of possible messages from the extremely broad space explored earlier (i.e., the space of words that appear in familiar phrases) to a more limited space involving permutations of the numbers 0 through 9 and the operations of addition and subtraction. On the other hand, this task remains especially challenging, because mishearing even one word of the problem would make it difficult or impossible to answer the question correctly. So if subjects can succeed at this task, this would suggest that they can essentially understand 
every word of an adversarial speech command when tested under more constrained conditions.

\vspace{.5cm}
\subsection{Methods}
\subsubsection{Stimuli}

We generated 20 arithmetic problems using the procedure described above. The problems contained two digits (from 0-9) and one instance of either the addition or subtraction operation. The problems were generated so that each digit (0-9) was the correct solution to two problems: one addition problem and one subtraction problem. 

Similarly to Experiments 3--4, 3 files contained arithmetic problems spoken by an uncorrupted human voice, and served as catch trials to ensure that subjects were engaged in the task and paying attention.

\subsubsection{Procedure}

Subjects were told a modified version of the story from previous experiments: 

\begin{quote}
\textit{A robot has hidden simple math problems in some audio transmissions that we've recovered. However, these audio clips have been ``corrupted''. Most of the transmissions sound very strange and garbled; the robot's ``voice'' is very different than a human voice.}

[...]

\textit{We want you to help us by solving the math problems. On each trial, you will listen to a short audio clip. These clips contain simple addition or subtraction problems, containing two numbers from zero to nine. After you play the clip, you will be presented with two possible answers to the problem. Your job is just to select the correct answer. For example, the robot voice might say ``5 plus 3''. If that happens, you should select ``8''.}
\end{quote}

This experiment proceeded in a very similar way to Experiments 3--4. After the subject played the clip, one button showed the correct answer, and one showed an incorrect answer. These pairs of options were randomly generated for each subject, as was the order in which the clips were shown. 

Experiment 5 also contained Catch trials of the same form as Experiments 3--4, and used the same exclusion criteria.

Readers can experience this task for themselves at \url{https://perceptionresearch.org/adversarialSpeech/E5}.

\begin{figure*}[!h]
\begin{center}
\vspace{.2cm}
\includegraphics[width=.6\linewidth]{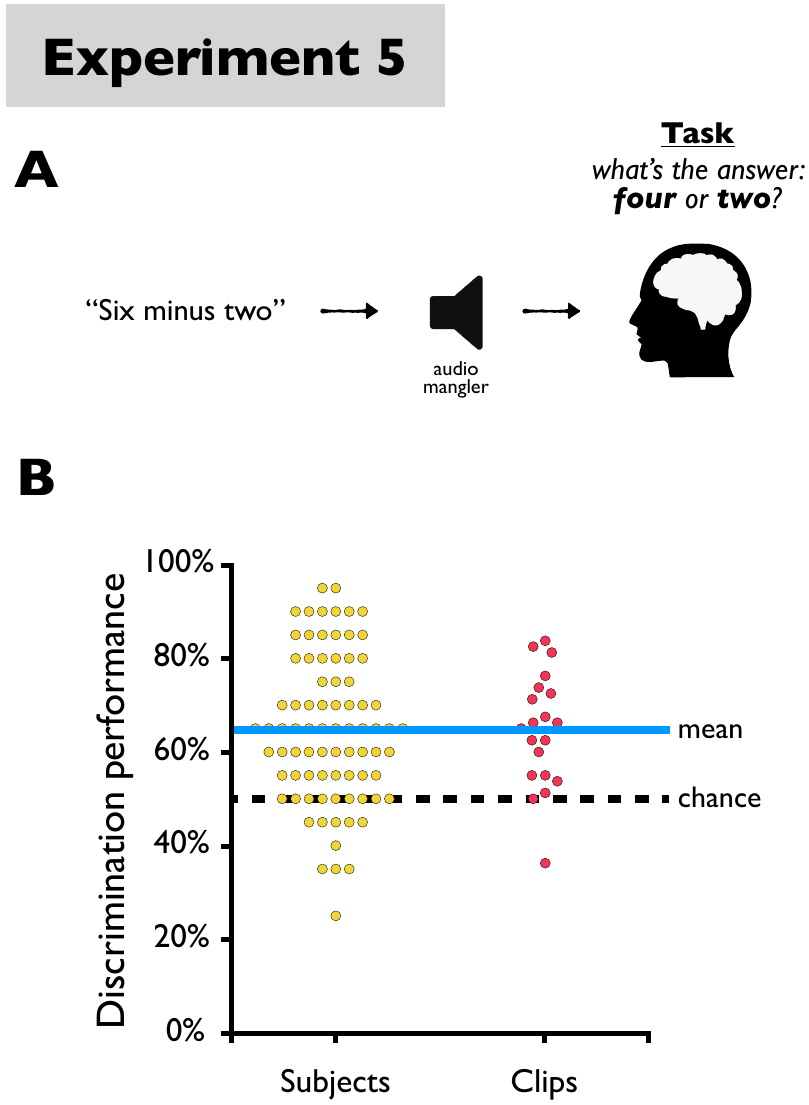}
\captionsetup{width=.9\linewidth,font=footnotesize}
\vspace{.1cm}
\caption
{\textit{Design and Results of Experiment 5}. (A) On each trial, subjects heard an adversarial speech clip expressing a simple arithmetic problem; their task was to supply the answer. (B) Subjects correctly answered the adversarial arithmetic problems, and this was again broadly true across both subjects (yellow) and speech clips (red).}
\label{fig:exp5}
\end{center}
\end{figure*}

\subsection{Results and Discussion}

Subjects demonstrated an ability to correctly answer arithmetic problems posed in adversarial speech. Average accuracy on experimental trials was 64.6\%, which significantly differed from chance (50\%), $t(79) = 8.27, p < .001, d = 0.930$ (Figure \ref{fig:exp5}B, blue). Thus, subjects could reliably comprehend most or all of the content of these audio commands, since such knowledge was necessary to answer the problem correctly. 

Moreover, collapsing across all speech clips, 81.9\% of subjects showed classification performance numerically above chance (Figure \ref{fig:exp5}B, yellow). Collapsing across all subjects, 18 out of 20 arithmetic problems were answered correctly more often than they were answered incorrectly (and 1 was answered correctly exactly half of the time) (Figure \ref{fig:exp5}B, red).

Whereas Experiments 3--4 could have been solved by understanding just one or two salient words, these results show that subjects displayed the ability to correctly decipher most (if not all) of the adversarial speech commands. 
In previous experiments, a subject who was played ``It is no use crying over spilled'' and noticed only the word ``spilled'' could still use that information to select ``milk'' as opposed to ``time''. However, here in Experiment 5, every word must be understood in order to solve these arithmetic problems, since even a single misheard word would make it nearly impossible to answer correctly. Thus, subjects can demonstrate an ability to understand whole phrases of adversarial speech, when tested in more sensitive ways.

One potential concern about this result is that it may not, after all, demonstrate comprehension of full speech commands, because subjects who heard only two words (e.g., ``7 plus ***'') or even just the operation itself (e.g., ``*** minus ***'') could still use this knowledge to perform above chance, even without comprehending the whole arithmetic problem. For example, consider a trial in which ``9 minus 5'' was played, but you heard only ``*** minus ***''; if you then noticed that the two options for that trial were ``9'' and ``4'', you might make the educated guess that ``4'' is more likely than ``9'' to be the correct answer, because there is only a single problem that could have the word ``minus'' in it while still having the answer ``9'' (i.e., the problem ``9 minus 0''), whereas many more problems containing ``minus'' could have the answer ``4'' (including, e.g., ``9 minus 5'', ``8 minus 4'', ``7 minus 3'', and so on). Since we have shown only that subjects perform above chance (and not, e.g., that they perform 
perfectly), it is possible that this above-chance performance merely reflects strategic responding based on partial knowledge, rather than complete understanding of adversarial speech commands.

However, this concern can be overcome by examining only those trials in which the correct answer was the 
less probable one given the logic above. For example, suppose ``9 minus 0'' was played, and the options were again ``9'' and ``4''. If subjects on such trials correctly answer ``9'' at rates above chance, then that above-chance performance could not be explained by strategic responding after hearing ``*** minus ***'' (or even ``9 minus ***''), since the optimal strategy in such cases of partial hearing would be 
not to answer ``9''. More generally, consider all trials in which either (a) the problem included ``plus'' and the correct answer was the 
lesser of the two options, or (b) the problem included ``minus'' and the correct answer was the 
greater of the two options. These trials are ones in which hearing only ``plus'' or ``minus'' would lead you 
away from the correct answer --- and so above-chance performance on such trials could not be explained by such strategic responding.

In fact, when we carried out this analysis, it revealed that even on those trials in which strategic responding based on partial knowledge would produce the incorrect answer, subjects still performed far above chance: 67.8\%, $t(79) = 8.53, p < .001, d = 0.960$. This analysis provides especially strong evidence that successful performance on math problems posed in adversarial speech goes beyond mere strategic responding, and implies a fuller and more complete understanding of the messages contained in such audio signals.

\vspace{.5cm}
\begin{center}
\section{General Discussion}
\end{center}
\vspace{.2cm}

What does it take to demonstrate that a human 
does not perceive a stimulus in a way that a machine 
does? Whereas previous work had identified a class of stimuli that machines comprehend but humans reportedly do not, here we showed that human subjects could display quite reliable and sophisticated understanding when tested in more sensitive ways. By taking adversarial speech as a case study, we showed that even when humans could not easily transcribe an adversarial speech stimulus (the previous benchmark for human understanding), they nevertheless 
could discriminate adversarial speech from closely matched non-speech (Experiments 1--2), finish common phrases started in adversarial speech (Experiment 3--4), and solve simple math problems posed in adversarial speech (Experiment 5) --- even though such stimuli have been previously described as ``unintelligible to human listeners''.

Of course, this work covers only one approach to generating adversarial speech (see below for discussion of other approaches), and adversarial speech is itself only one example of apparent divergence between human and machine perception. Nevertheless, these results show how 
sensitive tests can reveal perceptual and cognitive abilities that were previously obfuscated by relatively ``insensitive'' tests, and in ways that directly inform comparisons between human and machine perception.
\vspace{.5cm}
\subsection{Increasing sensitivity}

In Experiments 1--2, subjects reliably discriminated adversarial speech clips from the same clips played backwards. Whereas previous work suggested that subjects ``believed our audio was not speech'', our tests showed that subjects not only can identify such clips as speech, but can do so even when compared to audio signals that are matched on numerous 
low-level properties (since they were just the very same clips played in reverse). At the very least, these experiments suggests that subjects can attend to the phonetic or phonological cues that minimally distinguish speech from non-speech, even when those cues are obscured by the adversarial-speech-generation process.

Experiments 3--4 showed that subjects can not only hear adversarial speech as speech, but can also comprehend the 
content of that speech. Whereas a subject could succeed in Experiments 1--2 simply by picking up patterns of prosody or segmentation, Experiments 3--4 asked subjects to fill in the last word of a phrase, and so required at least some 
comprehension of the adversarial speech clip. Moreover, this result could not be explained by straightforward forms of ``priming''. For example, \cite{Carlini:vl} rightly note that, if one is told in advance what to hear in an adversarial speech clip, it is surprisingly easy to hear that message. But in our Experiments 3--4, the subjects were given no advance knowledge about which words they would encounter. Indeed, even though subjects knew that the phrases would likely be familiar, simply being told that one will hear a familiar phrase says little about 
which words will appear in that phrase. Instead, it tells you only that, 
if you can make out one or two salient words (e.g., ``picture'' and ``thousand''), you might be likely to correctly guess what will come next (e.g., ``words''). But even this situation first requires understanding some aspects of the speech clip, and so demonstrates a kind of comprehension that previous studies failed to reveal.

Finally, Experiment 5 created conditions that demonstrate complete (or nearly complete) comprehension of adversarial speech, by playing subjects spoken arithmetic problems and asking them to select the answer. For these problems (e.g., ``six minus two''), every word of the problem was crucial to completing the task, since even a single misheard word would undermine one's ability to answer it. In a sense, this task is nearly equivalent to the free transcription task from previous work (since it required near-perfect comprehension to succeed), except that this experiment involved a more constrained space of word possibilities (since the subjects knew to expect the numbers 0 through 9, and the words ``plus'' and ``minus''). Nevertheless, there was still considerable uncertainty for any message: Even under these constraints, there were 110 possible problems that could have appeared on any given trial, such that subjects did not have a straightforward or trivial way to know what they were ``supposed to hear'' in the message. 

\vspace{.2cm}
\subsection{Psychophysically inspired}

Our approach here was motivated by a central insight from the human psychophysics literature: simply asking people to describe what they see, hear, feel, or know can severely underestimate the nature and extent of such knowledge. Of the many reasons for this, one is that subjects may adopt conservative or idiosyncratic criteria when generating such explicit reports. We perceptually experience more than we can hope to include in any report of such experiences, and so when faced with very unconstrained tasks (such as freely describing what we hear in an audio clip), we must make choices about which aspects of our experience to describe, and in how much detail. These choices will be influenced by a host of factors, including which aspects of our experience stand out as remarkable, which we think the experimenter wants to hear about, and even just how engaged and motivated we are by the task. By contrast, more constrained tasks of the sort we explore here have the ability to zero in on those aspects of the subject's knowledge we are interested in, and in ways that require the subject to make very few (if any) choices about the relevance of various kinds of information. Indeed, similar considerations may even apply to tests of 
machine perceptual knowledge itself. For relevant work on this issue, see \citealp{Zoran_2015_ICCV,Ritter:2017uk}.

In fact, strictly speaking, what we have explored here is not literally a task that is somehow more  sensitive, but rather a task that makes adequate a previously inadequate 
measure. Whereas previous work adopted ``\% correct'' as the standard for comprehension of adversarial speech, this measure turns out to have been inadequate to reveal such comprehension when applied to a free transcription task. Perhaps some other measure (e.g., more advanced text mining) could, when applied to free transcription, capture the higher levels of comprehension that we observe here. But for our two-alternative forced-choice and forced-response tasks, the ``\% correct'' measure was indeed adequate, because the task constrained subjects' responses to the variables and dimensions of interest, thereby revealing knowledge that was hidden by previous approaches.

The present work also joins other projects that have begun to explore similar themes. For example, \cite{vadillo2020human} use adversarial speech stimuli consisting of regular audio overlaid with specific noise patterns that cause a speech recognition system to misclassify the audio clips. These authors also advocate for careful human subjects testing before describing an adversarial audio attack as ``imperceptible'', though their studies primarily focus on the more basic capacity to distinguish clips containing adversarial attacks from normal clips. By contrast, our present studies reveal that, in at least some cases, subjects can not only identify that adversarial clips contain speech, but can also comprehend the content of some adversarial audio attacks\footnote{We note that \citet{vadillo2020human} also demonstrate that subjects found some adversarial stimuli to be significantly less natural non-adversarial stimuli. This difference is more in line with the present work, as it demonstrates subjects' perception of specific differences between clips that may signal that the stimuli contains an adversarial attack.}. Other work has investigated human vs machine perception of ``deepfake'' stimuli, including both auditory and visual examples \citep{groh2022deepfake,muller2021human}.  

\vspace{.2cm}
\subsection{Consequences for human-machine comparisons}

Increased understanding of the sort revealed here matters for at least two kinds of reasons:

First, and more practically, a major source of interest in adversarial attacks and other machine misclassifications derives from the security concerns they raise about the vulnerability of various machine-learning applications. For example, adversarial images could be used to fool autonomous vehicles into misreading street signs, and adversarial speech could be used to attack smartphones and home assistants without a human supervisor knowing. But the present results suggest that, at least in some circumstances, humans may be more aware of such attacks than was previously evident. Indeed, at least for the attack explored here, our Experiments 1--2 suggests that humans may well know 
\textit{that} they are being attacked, even without knowing the precise 
way in which they are being attacked. And our Experiments 3--5 suggest that they may even have more sophisticated knowledge of the content of such attacks. Indeed, the constraint in Experiment 5, involving numbers (``0'', ``1'', ``2'', …, ``9'') and a few keywords (``plus'', ``minus'') is closely analogous to certain kinds of attacks that malicious actors might actually deliver to a phone (e.g., ``dial 911''). 

But second, and more theoretically, human understanding of stimuli that fool machines is of interest to work in cognitive science (including both psychology and artificial intelligence) that explores similarities and differences between human perception and the analogous processes in various machine-learning systems. Adversarial attacks in particular are frequently invoked as evidence of ``an astonishing difference in the information processing of humans and machines'' (\citealp{brendel2020adversarial}; see also \citealp{HendrycksG16b,Sabour:2015vd,Serre:2019cm,buckner2019comparative,ilyas2019adversarial}). But while it seems clear that humans and machines don't perceive such stimuli 
identically, it is still of interest to know just 
how similar or different their perception of such stimuli is. As we show here, the answer to this question can be surprisingly subtle: Humans who cannot freely report the content of such stimuli may nevertheless decipher them under certain conditions, in ways directly relevant to claims about overlapping or non-overlapping processing across such systems.

Of course, none of the present experiments imply that 
all adversarial attacks will be comprehensible to humans in this way (though there are indeed other audio adversarial attacks that subjectively sound speechlike; \citealp{Abdullah:2019uh}). For example, some audio adversarial attacks involve ultrasonic frequencies beyond the range of human hearing \citep{Zhang:2017kl}; these attacks will certainly be incomprehensible to people. At the same time, that very fact can make them ``weaker'' or less threatening as attacks. For example, the developers of this attack note that an automated speech-recognition system could ``defend'' against it by restricting its processing to the frequency bands of human hearing (e.g., by modifying a microphone to ``suppress any acoustic signals whose frequencies are in the ultrasound range''). And more generally, if the indecipherability of such attacks to humans owes fully to their occurring outside the range of human auditory perception, then this too is less an example of deep underlying differences in speech processing and more so an example of superficially different ``performance constraints'' 
facing these two systems \citep{firestone2020performance}.

Relatedly, it has also been shown that adversarial speech can be embedded in clips of normal human speech \citep{carlini:v2, Qin:2019vz}. These attacks seem particularly difficult to decipher, but at the same time the original authors note here too that they are audible if you ``listen closely'' \citep{carlini:online}, or more generally that they ``can still be differentiated from the clean audio'', in such a way that could still make them detectable 
as attacks. All these cases, then, show just how much nuance is required to make valid comparisons across human and machine perception. 

\vspace{.2cm}
\subsection{Broader lessons}

Though the present experiments explore sensitive comparisons in a case study involving adversarial speech, this approach could apply much more broadly to nearly any comparison of human and machine perception. Indeed, even recent work that does not use the language of ``sensitive tests'' may still be considered within this framework. For example, it had previously been claimed that it is possible to generate bizarre visual images that machines recognize as familiar objects but which are ``totally unrecognizable to human eyes'' \citep{Nguyen:2015cv}. However, follow-up work using forced-responding showed that human observers can actually anticipate machine classifications when given relevant alternatives to choose from (\citealp{Zhou:2019jxa}; though see \citealp{dujmovic2020adversarial}). For example, a series of cross-hatched red lines on a white background is recognized as a ``baseball'' by AlexNet; even though a human may not have initially been inclined to give the image that label, they are easily able to \textit{select} ``baseball'' over relevant alternatives under forced-response conditions not unlike those we explore here. 

More generally, the approach we advocate here 
could apply to much broader questions about the overlap of human and machine processing. 
Even beyond adversarial misclassification, recent work has shown that natural language inference systems often rely on surface-level heuristics to make judgements about whether one sentence logically entails another \citep{mccoy2019right}. This work also includes a human-machine comparison, where it is concluded that ``human errors are unlikely to be driven by the heuristics targeted'' in that work. This work could also benefit from a more sensitive test, such as one that eliminates the ability to reread the sentences, or perhaps introduces a time constraint. Under these strict constraints, it is possible that human judgments would be more informed by surface-level heuristics, revealing that some aspect of human cognition are reflected in the mistakes of the natural language inference systems. 

Another human-machine difference that could benefit from more sensitive tests is the finding that Deep Convolutional Neural Nets tend to classify images based on texture rather than shape \citep{Baker:2018jp}, whereas human subjects tend to classify based on shape rather than texture \citep{Landau:1988gs}. But the human subjects in \cite{Baker:2018jp} were almost completely unconstrained in their testing conditions, being able to view the images and consider their labels for as long as they like. Perhaps forcing subjects to classify quickly, and after only a brief presentation, could bring the human and machine classification judgments into better alignment (see also \citealp{geirhos2018imagenet}). 
For an exploration of this possibility, see \cite{hermann2022understanding}. Whereas the present work has used sensitive tests to reveal machine-like capabilities on a difficult task, these two examples demonstrate ways in which sensitive testing can be used to reveal machine-like deficiencies on fairly straightforward tasks.

Finally, even though the example we explore here shows how sensitive tests can reveal 
similarities where there previously seemed to be 
dissimilarities, the opposite pattern of results is possible as well. First, for some future class of stimuli, sensitive tests could well reveal that humans 
cannot comprehend, perceive, or process them the way a machine does. In that case, researchers could become especially confident in a given human-machine difference that survives even a sensitive test. Second, sensitive tests could also isolate very specific differences in how a human and a machine classify a stimulus. For example, if a human views the ``baseball'' image from \cite{Nguyen:2015cv} and consistently prefers a specific alternative label (e.g., ``chainlink fence''), this would suggest all the more strongly that the two systems represent this image differently.

\vspace{.2cm}
\subsection{Concluding Remarks}

People know and experience more than they freely report. Though this is a familiar and often-studied problem in cognitive psychology and perception research, it is also relevant to research 
comparing human perception and cognition to the analogous processes in machines. Here, we have shown how lessons from human perception research can directly inform and advance such comparisons, including in ways that reveal latent or implicit knowledge that was not evident from initial (and perhaps insensitive) comparisons. We thus advocate the adoption of more sensitive tests of human and machine perception, so that we can better explore when humans and machines do --- or do not --- perceive the world the same way.

\section{Acknowledgments}


For helpful discussion and/or comments on previous drafts, we thank Tom McCoy, Ian Phillips, and members of the JHU Perception \& Mind Lab. For assistance with beeswarm plots, we thank Stefan Uddenberg. For resources relating to the production of adversarial speech commands, we thank Nicholas Carlini. This work was supported by a JHU ASPIRE Grant (M.L.) and the JHU Science of Learning Institute (C.F.).

\section{Author Contributions}

M.L. and C.F. designed the experiments and wrote the paper. M.L. ran the experiments and analyzed the data, under the supervision of C.F.

\section{Data Availability}

The data, materials, code, and pre-registrations supporting all of the above experiments are available at
\url{https://osf.io/kp5j9/}.

\bibliographystyle{apalike}
\bibliography{reference}




\appendix

\section{Replicating Experiment~1 with original adversarial speech clips}
\label{carliniStimRep}
All our experiments generated new stimuli from scratch using the whitebox method described by \citet{Carlini:vl}. However, our methods apply equally well to other stimuli generated using this method, including those presented in the original work. In a preregistered supplementary study, we adopted the same procedure from Experiment~1, except we use the 5 stimuli provided at \url{https://www.hiddenvoicecommands.com/white-box}. We converted those files to .wav format, trimmed them to have a shorter period of silence following the command, amplified them, and then played them either forwards or backwards. 100 subjects (before exclusions) were recruited from Prolific (see \citealp{peer2017beyond}). We included one catch trial of the same form as Experiment~1. 83 subjects were included in the analysis after exclusions. The mean accuracy on experimental trials was 68.9\%, which significantly differed from chance accuracy (50\%), $t(82) = 8.20, p < .001, d = 0.906$. These results demonstrate that subjects can also reliably determine whether the audio attacks presented in \citet{Carlini:vl} contain speech, consistent with our findings in Experiment~1.
\end{document}